\documentclass[10pt]{iopart}
\usepackage{iopams}         
\usepackage[english]{babel} % Change hyphenation rules
\usepackage{graphicx} % Add pictures to your document
\usepackage{listings} % Source code formatting and highlighting
\usepackage{soul}
\usepackage{color}
\usepackage{lipsum}
\usepackage{float}
\usepackage{amsmath}

\usepackage{comment}

\begin{document}

\title[Charge affinity and solvent effects in numerical simulations of ionic microgels]{Charge affinity and solvent effects in numerical simulations of ionic microgels}

\author{Giovanni Del Monte$^{1,2,3,*}$, Fabrizio Camerin$^{1,4,*}$, Andrea Ninarello$^{1,2}$, Nicoletta Gnan$^{1,2}$, Lorenzo Rovigatti$^{2,1}$, Emanuela Zaccarelli$^{1,2,*}$}
\address{$^1$ CNR Institute of Complex Systems, Uos Sapienza, piazzale Aldo Moro 2, 00185, Roma, Italy}
\address{$^2$ Department of Physics, Sapienza University of Rome, piazzale Aldo Moro 2, 00185 Roma, Italy}
\address{$^3$ Center for Life NanoScience, Istituto Italiano di Tecnologia, viale Regina Elena 291, 00161 Rome, Italy}
\address{$^4$ Department of Basic and Applied Sciences for Engineering, via Antonio Scarpa 14, 00161 Roma, Italy}
\ead{giovanni.delmonte@uniroma1.it}
\ead{fabrizio.camerin@uniroma1.it}
\ead{emanuela.zaccarelli@cnr.it}

\vspace{10pt}
\begin{indented}
\item[]\today
\end{indented}

\begin{abstract}
	Ionic microgel particles are intriguing systems in which the properties of thermo-responsive polymeric colloids are enriched by the presence of charged groups. In order to rationalize their properties and predict the behaviour of microgel suspensions, it is necessary to develop a coarse-graining strategy that starts from the accurate modelling of single particles. Here, we provide a numerical advancement of a recently-introduced model for charged co-polymerized microgels by improving the treatment of ionic groups in the polymer network. We investigate the thermoresponsive properties of the particles, in particular their swelling behaviour and structure, finding that, when charged groups are considered to be hydrophilic at all temperatures, highly charged microgels do not achieve a fully collapsed state, in favorable comparison to experiments. In addition, we explicitly include the solvent in the description and put forward a mapping between the solvophobic potential in the absence of the solvent and the monomer-solvent interactions in its presence, which is found to work very accurately for any charge fraction of the microgel.
	Our work paves the way for comparing single-particle properties and swelling behaviour of ionic microgels to experiments and to tackle the study of these charged soft particles at a liquid-liquid interface.
\end{abstract}

\vspace{2pc}

\vspace{2pc}
\noindent\submitto{\JPCM}
\ioptwocol

\section{Introduction}

Soft matter is a very active branch of condensed matter physics, which comprises, among other systems, colloidal suspensions, whose constituent particles can greatly vary in shape, softness and function. Soft matter encompasses not only synthetic particles, but also constituents of many biological systems, such as proteins, viruses and even cells, whose size ranges between the nano and the micrometer scale. 
A peculiar aspect of soft matter systems is the great variety of amorphous states they can form, including glasses~\cite{sciortino2005glassy,pusey2008colloidal} and gels~\cite{zaccarelli2007colloidal,lu2013colloidal,joshi2014dynamics}. Indeed, a large amount of work in this field is devoted to the study of these non-ergodic states which may form due to different kind of interactions, such as steric, hydrophobic or electrostatic ones, both of attractive and repulsive nature.

Sometimes, a single colloidal particle is already quite a complex object whose behaviour at the collective level is strongly connected to the microscopic features of the particle itself. This situation is typical of soft colloids, i.e. deformable particles with internal degrees of freedom strongly influencing their mutual interactions, which makes them already intrinsically multi-scale. For these systems a theoretical approach is quite challenging even at the single-particle level, thus it is convenient to rely on the development of suitable coarse-grained models~\cite{likos2001effective} that allow to greatly reduce the system complexity, still capturing the important ingredients to be retained for a correct description of the collective behavior. This strategy is very profitable for the case of microgel particles~\cite{fernandez2011microgel} that, combining together the properties of colloids and polymers, can be viewed as a prototype example of soft particles
~\cite{lyon2012polymer,vlassopoulos2014tunable}. A microgel is a microscale gel whose internal polymeric network controls its peculiar properties. By varying the constituent monomers, microgels can be made responsive to temperature, pH or to external forces~\cite{fernandez2011microgel}. For their intriguing properties, they are employed in a wide variety of applications, ranging from biomedical purposes~\cite{oh2008development,karg2019nanogels} to paper restoration~\cite{di2020gellan}.

In order to be able to predict the  behaviour of dense microgel suspensions and the formation of arrested states, it is important to properly take into account the internal degrees of freedom of the particles, by modelling their effective interactions in such a way that the resulting object can still shrink, deform and interpenetrate~\cite{rovigatti2019connecting,bergman2018new,conley2019relationship,gnan2019microscopic}. Hence, an accurate modelling of a single microgel is a necessary pre-requisite for a correct description of bulk suspensions.  To validate numerical models at the single-particle level, there are a number of different experiments we can refer to. One of the most straightforward is the measurement of the effective size of the microgels via dynamic light scattering experiments. Upon varying the controlling parameter of the dispersion, the so-called swelling curves can be determined. For instance, microgels synthesized by employing a thermoresponsive polymer, such as Poly(N-isopropyl-acrylamide) (PNIPAM), undergo a so-called Volume Phase Transition (VPT)~\cite{fernandez2011microgel} at a temperature $T_{\scriptscriptstyle \mathit{VPT}} \approx 32^{\circ}$C from a swollen to a collapsed state.
\\
In addition, form factors can be measured by small-angle scattering experiments of dilute microgel suspensions, either using neutrons~\cite{stieger2004small}, x-rays~\cite{ninarello2019modeling} or even visible light for large enough microgels~\cite{bergman2020controlling}. This observable directly provides information on the inner structure of the microgels and  shows that microgels prepared via precipitation polymerization~\cite{pelton2011microgels} can be modelled as effective fuzzy spheres~\cite{stieger2004small}, where a rather homogeneous core is surrounded by a fluffy corona, giving rise to what is usually called a core-corona structure.
A more complex situation arises when ionic groups are added to the synthesis to make microgels responsive also to external electric fields~\cite{nojd2013electric,colla2018self} and to pH variations~\cite{nigro2015dynamic}.
 A case study of such these co-polymerized microgels is the one made of PNIPAM and polyacrylic acid (PAAc)~\cite{pelton2011microgels,nojd2018deswelling,rochette2017effect,capriles2008coupled}, that is pH-responsive due to the the weak acidic nature of AAc monomers.

An increasing amount of work in the last years has focused on modelling single-particle behaviour both of neutral~\cite{gnan2017silico,moreno2018computational,rovigatti2019numerical} and ionic microgels~\cite{quesada2012computer,kobayashi2014structure,martin2019review,del2019numerical}. For the latter case, we have recently shown~\cite{del2019numerical} that it is important  to take into account both the disordered nature of the network, as opposed to the diamond lattice structure~\cite{rovigatti2019numerical}, and an explicit treatment of charges and counterions. Indeed, mean-field approaches completely neglect the complex, heterogeneous nature of the charge distribution within the microgel.

In this work, we extend our previous effort by going one step further towards a realistic numerical treatment of ionic co-polymerized microgels.
In Ref.~\cite{del2019numerical}, we modelled a single microgel particle such that all of its monomers, including charged ones, experienced a solvophobic attraction on increasing temperature. Here, instead, charged monomers experience Coulomb and steric interactions only. This is expected to be more realistic, since charged or polar groups always remain hydrophilic irrespectively of temperature, thus having a distinct behaviour with respect to all other monomers. We examine the consequences of this difference on the microgel swelling behaviour as well as on its structure and charge distributions across the VPT. In the second part of the manuscript, we consider the presence of an explicit solvent, to examine its effects on the structural properties of the microgel. In this way, we aim to make our model suitable for situations where solvent effects become important. In particular, this will enable us to study the effect of charges for microgels adsorbed at liquid-liquid interfaces, similarly to what we recently put forward for neutral microgels~\cite{camerin2019microgels,camerin2020microgels}. 

\section{Methods}
\small
The coarse-grained microgels used in this work are prepared as in Refs.~\cite{gnan2017silico} starting from $N$ patchy particles of diameter $\sigma$, which sets the unit of length, confined 
in a spherical cavity. A fraction $c=0.05$ of these particles has four patches on their surface to mimic the typical crosslinker connectivity in a chemical synthesis, while all the others have two patches to represent monomers in a polymer chain. During the assembly, an additional force is employed on the crosslinking particles to increase their concentration in the core of the microgel in agreement with experimental features~\cite{ninarello2019modeling}. Once a fully-bonded configuration 
is reached (when the fraction of formed bonds is greater than $0.995$), a permanent topology is obtained by enforcing the Kremer-Grest bead-spring model~\cite{kremer1990dynamics}. In this way, all particles experience a steric repulsion via the Weeks-Chandler-Anderson (WCA) potential,
\begin{equation}
V_{\rm WCA}(r)=
\begin{cases}
4\epsilon\left[\left(\frac{\sigma}{r}\right)^{12}-\left(\frac{\sigma}{r}\right)^{6}\right] + \epsilon & \text{if $r \le 2^{\frac{1}{6}}\sigma$}\\
0 & \text{otherwise,}
\end{cases}
\end{equation}
where $\epsilon$ sets the energy scale and $r$ is the distance between the centers of two beads.
Additionally, bonded particles interact via the Finitely Extensible Nonlinear Elastic potential (FENE),
\begin{equation}
V_{\rm FENE}(r)=-\epsilon k_FR_0^2\ln\left[1-\left(\frac{r}{R_0\sigma}\right)^2\right]     \text{ if $r < R_0\sigma$,}
\end{equation}
with $k_F=15$ which determines the stiffness of the bond and $R_0=1.5$ which determines the maximum bond distance.  
The resulting microgel is thus constituted by a disordered polymer network with a core-corona structure and form factors across the VPT that closely resemble experimental ones for microgels synthesized via the precipitation polymerization procedure~\cite{ninarello2019modeling}. 
\\
Once the microgels are prepared, we add electrostatic interactions to mimic co-polymerized polymer networks with charged groups. To this aim, we randomly assign a negative charge $-e^*$ to a given fraction $f$ of microgel monomers, to mimic the acrylic acid dissociation in water, where $e^*=\sqrt{4\pi\varepsilon_0\varepsilon_r\sigma\epsilon}$ is the reduced unit charge (which roughly corresponds to the elementary charge $e$, considering $\epsilon\approx k_BT$ at room temperature and $\sigma$ as the polymer's Kuhn length), and $\varepsilon_0$ and $\varepsilon_r$ are the vacuum and relative dielectric constants.
Accordingly, we insert in the simulation box an equal number of positively charged counterions with charge $e^*$ to ensure the neutrality of the system. 
Interactions among charged beads are given by the reduced Coulomb potential
\begin{equation}
V_{\rm coul}(r)=\frac{q_i q_j \sigma}{e^{*2} r}\epsilon,
\end{equation}
where $q_i$ and  $q_j$ are the charges of counterions or charged monomers. We adopt the particle-particle-particle-mesh method~\cite{deserno1998mesh} as a long-range solver for the Coulomb interactions. As discussed in a previous contribution~\cite{del2019numerical}, the size of the counterions is set to $0.1\sigma$ to facilitate their diffusion within the microgel network and to avoid spurious excluded volume effects. They interact with all other species simply via the WCA potential.
\\
The swelling behaviour of a thermoresponsive microgel can be reproduced in molecular dynamics simulations either by means of an implicit solvent, namely by adding a potential that mimics the effect of the temperature on the polymer, or by explicitly adding coarse-grained solvent particles within the box. In the first case, we employ a solvophobic potential
\begin{equation}\label{eq:valpha}
V_{\alpha}(r)  =  
\begin{cases}
-\epsilon\alpha & \text{if } r \le 2^{1/6}\sigma  \\
\frac{1}{2}\alpha\epsilon\left\{\cos\left[\gamma{\left(\frac{r}{\sigma}\right)}^2+\beta\right]-1\right\} & \text{if } 2^{1/6}\sigma < r \le R_0\sigma  \\
0 & \text{if } r > R_0\sigma
\end{cases}
\end{equation}
with $ \gamma = \pi \left(\frac{9}{4}-2^{1/3}\right)^{-1} $ and $\beta = 2\pi - \frac{9}{4}\gamma$~\cite{soddemann2001generic}. This potential introduces an effective attraction among polymer beads, modulated by the parameter $\alpha$, whose increase corresponds to an increase in the temperature of the dispersion. For $\alpha=0$ no attraction is present, which corresponds to fully swollen, i.e. low-temperature, conditions. For neutral microgels, the VPT is encountered at $\alpha \approx 0.65$, while a full collapse is usually reached for $\alpha \gtrsim 1.2$.

In the first part of this work, we analyze two different models, based on a different treatment of the interactions of charged ions on the microgels: in \textit{Model I}  all monomers experience a total interaction potential where $V_\alpha$ (Eq.~\ref{eq:valpha}) is added to the Kremer-Grest interactions, as previously done in Ref.~\cite{del2019numerical}; in \textit{Model II} only neutral monomers experience this total interaction potential, while the charged monomers do not interact with $V_\alpha$, i.e. $\alpha=0$ for them in all cases.  This second situation is equivalent to leaving unaltered the behaviour of charged groups of the microgel as the solvent conditions change, so that they always retain a good affinity for the solvent ($\alpha=0$). A similar treatment is also adopted for the counterions, for which $\alpha=0$ for both \textit{Model I} and \textit{Model II}.
\\
\\
In the second part of this work, we explicitly consider the presence of the solvent in driving the Volume Phase Transition. Solvent particles are modelled within the Dissipative Particle Dynamics (DPD) framework in order to avoid spurious effects which may arise from the use of a standard Lennard-Jones potential due to the excessive excluded volume of the solvent~\cite{camerin2018modelling}. In the DPD scheme, two particles $i$ and $j$ experience a force $\vec{F}_{ij} = \vec{F}^C_{ij} + \vec{F}^D_{ij} + \vec{F}^R_{ij}$, where:
\begin{eqnarray}
	\vec{F}^C_{ij}  &=&  a_{ij} w(r_{ij}) \hat{r}_{ij} \\
	\vec{F}^D_{ij}  &=&  -\gamma w^2(r_{ij}) (\vec{v}_{ij}\cdot\vec{r}_{ij}) \hat{r}_{ij} \\
	\vec{F}^R_{ij}  &=&  2\gamma\frac{k_B T}{m} w(r_{ij}) \frac{\theta}{\sqrt{\Delta t}} \hat{r}_{ij}
\end{eqnarray}
where $\vec{F}^C_{ij}$ is a conservative repulsive force, with $w(r_{ij}) = 1-r_{ij}/r_c$ for $r_{ij}<r_c$ and $0$ elsewhere, $\vec{F}^D_{ij}$ and $\vec{F}^R_{ij}$ are a dissipative and a random contribution of the DPD, respectively; $\gamma$ is a friction coefficient, $\theta$ is a Gaussian random variable with average $0$ and unit variance, and $\Delta t$ is the integration time-step. We set $r_c = 1.75\sigma$ and $\gamma = 2.0$ in all the simulations. Here $a_{i,j}$ quantifies the repulsion between two particles $i$ and $j$, which effectively allows the tuning of the monomer-solvent (m,s) and solvent-solvent (s,s) interactions. Following our previous work~\cite{camerin2018modelling}, we fix $a_{s,s}=25.0$ while we vary $a_{m,s}\equiv a$ between 5.0 and 16.0, that is the range where the collapse of a neutral microgel takes place. The reduced DPD density is set to $\rho_s r_c^3 = 3.9$ (with $\rho_s$ the actual number density of solvent beads).
With this choice of parameters, we previously showed that this model reproduces the swelling behaviour  and structural properties of a neutral microgel particle, in quantitative agreement with the implicit solvent model that was explicitly tested against experiments~\cite{ninarello2019modeling}. To compare the explicit solvent model with the implicit one, we only consider \textit{Model II}, where charged monomers always retain a high affinity for the solvent. We will show later that, in the explicit treatment, this translates to having charged monomers-solvent interactions (ch,s) always set to $a_{ch,s}=0$. All other interactions are identical to the implicit solvent model.
\\
Simulations are performed with LAMMPS~\cite{plimpton1995fast}. The equations of motion are integrated with the velocity-Verlet algorithm. All particles have unit mass $m$, the integration time-step is $\Delta t = 0.002 \sqrt{m\sigma^2/\epsilon}$ and the reduced temperature $T^*=k_BT/\epsilon$ is set to 1.0 by means of a Nos\`e-Hoover thermostat for implicit solvent simulations or via the DPD thermostat for explicit solvent ones. In the former case the number of monomers in the microgels is fixed to $N\approx 42000$, while for the latter case we limit the analysis to $N\approx 5000$ due to the very large computational times owing to the presence of the solvent.
\\
\\
From equilibrated trajectories, we directly calculate the form factor of the microgel as,
\begin{equation}
P(q)=\left\langle\frac{1}{N}\sum_{i,j=1}^N  \exp{(-i\vec{q} \cdot \vec{r}_{ij})} \right\rangle,
\end{equation} 
where $r_{ij}$ is the distance between monomers $i$ and $j$, while the angular brackets indicate an average over different configurations and over different orientations of the wavevector $\vec{q}$ (for each $q$ we consider $300$ distinct directions randomly chosen on a sphere of radius $q$).
\\
Also, we determine the radius of gyration $R_g$ of the microgels as a measure of their swelling degree. This is calculated as 
\begin{equation}
R_g=\left\langle\left[\frac{1}{N}\sum_{i=1}^N (\vec{r}_i-\vec{r}_{CM})^2\right]^{\frac{1}{2}} \right\rangle ,
\end{equation} 
where $\vec{r}_{CM}$ is the position of the center of mass of the microgel.
\\
For each swelling curve, representing $R_g$ as a function of the effective temperature $\alpha$ (implicit solvent) or $a$ (explicit solvent), we define an effective VPT  temperature, either $\alpha_{\scriptscriptstyle \mathit{VPT}}$ or $a_{\scriptscriptstyle \mathit{VPT}}$, as the inflection point of a cubic spline interpolating the simulation points.
\\
Finally, the radial density profile for all monomers is defined as 
\begin{equation}\label{eq:profile}
\rho(r)= \left\langle \frac{1}{N}\sum_{i_{=1}}^{N} \delta (|\vec{r}_{i}-\vec{r}_{CM}|-r) \right\rangle.
\end{equation} 
By restricting the sum in Eq.~\ref{eq:profile} to only charged monomers or to counterions, we also calculate $\rho_{CH}(r)$ and $\rho_{CI}(r)$, that are the radial density profiles of charged microgel monomers and of counterions, respectively. In addition, we define the net charge density profile as
\begin{equation}\label{eq:rho_charge}
\rho_Q(r) = -\rho_{CH}(r)+\rho_{CI}(r).
\end{equation}

\normalsize
\section{Results and Discussion}
\begin{figure*}[htbp]
	\centering
	\includegraphics[scale=0.5]{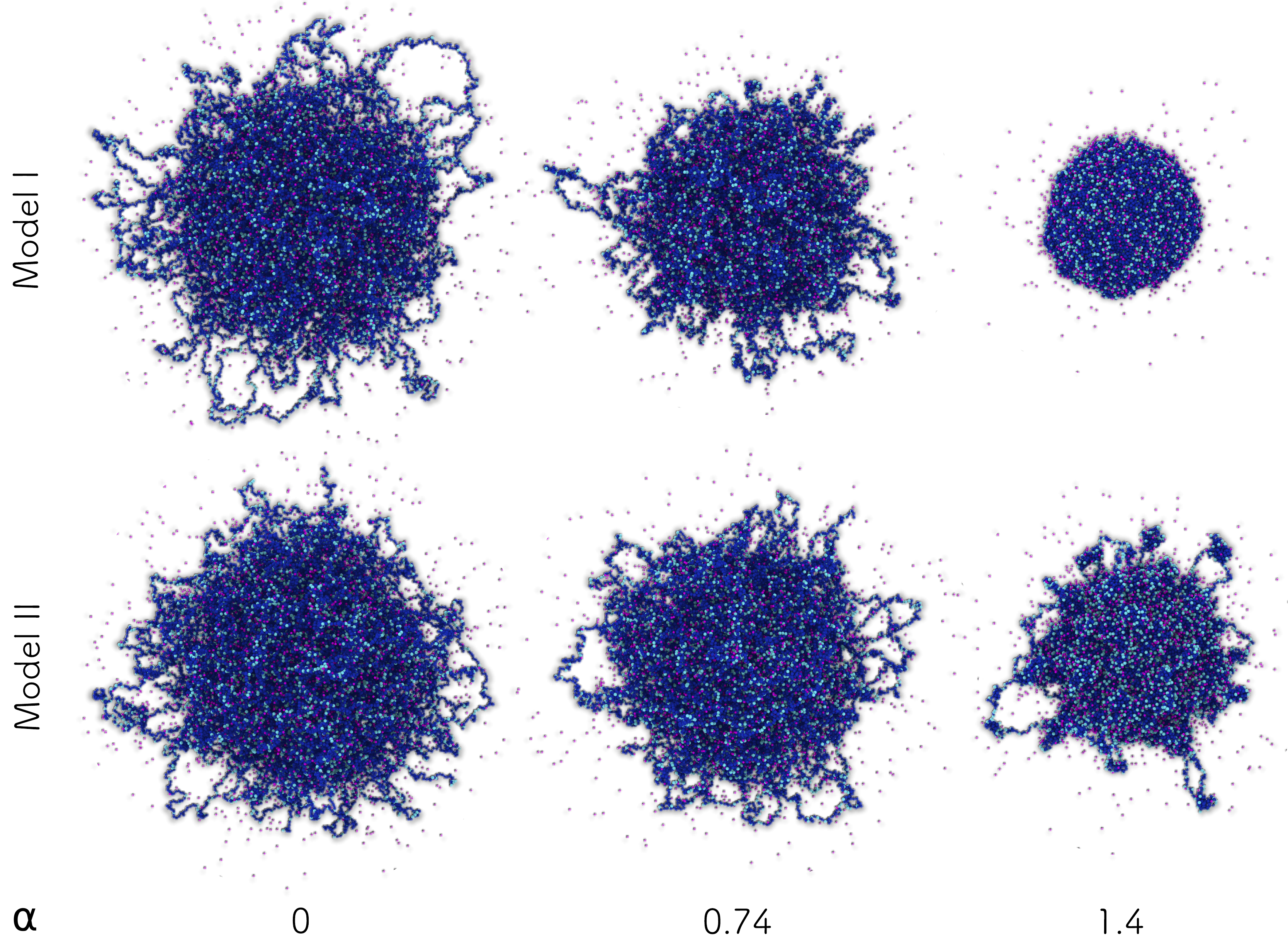}
	\caption{\textbf{Simulation snapshots.} Ionic microgels with $f=0.2$ and $N\approx 42000$ obtained in implicit solvent for (top) \textit{Model I} and (bottom) \textit{Model II} for $\alpha=0, 0.74$ and $1.4$ (from left to right panels). Blue (cyan) particles are neutral (charged) microgel monomers; smaller purple particles represent counterions. 
	}
	\label{fig:snaps42k}
\end{figure*}

\subsection{\textbf{On the role of the affinity of charged monomers for the solvent}}
\subsubsection{Swelling behaviour}

\begin{figure}[t!]
\centering
\includegraphics[scale=0.51]{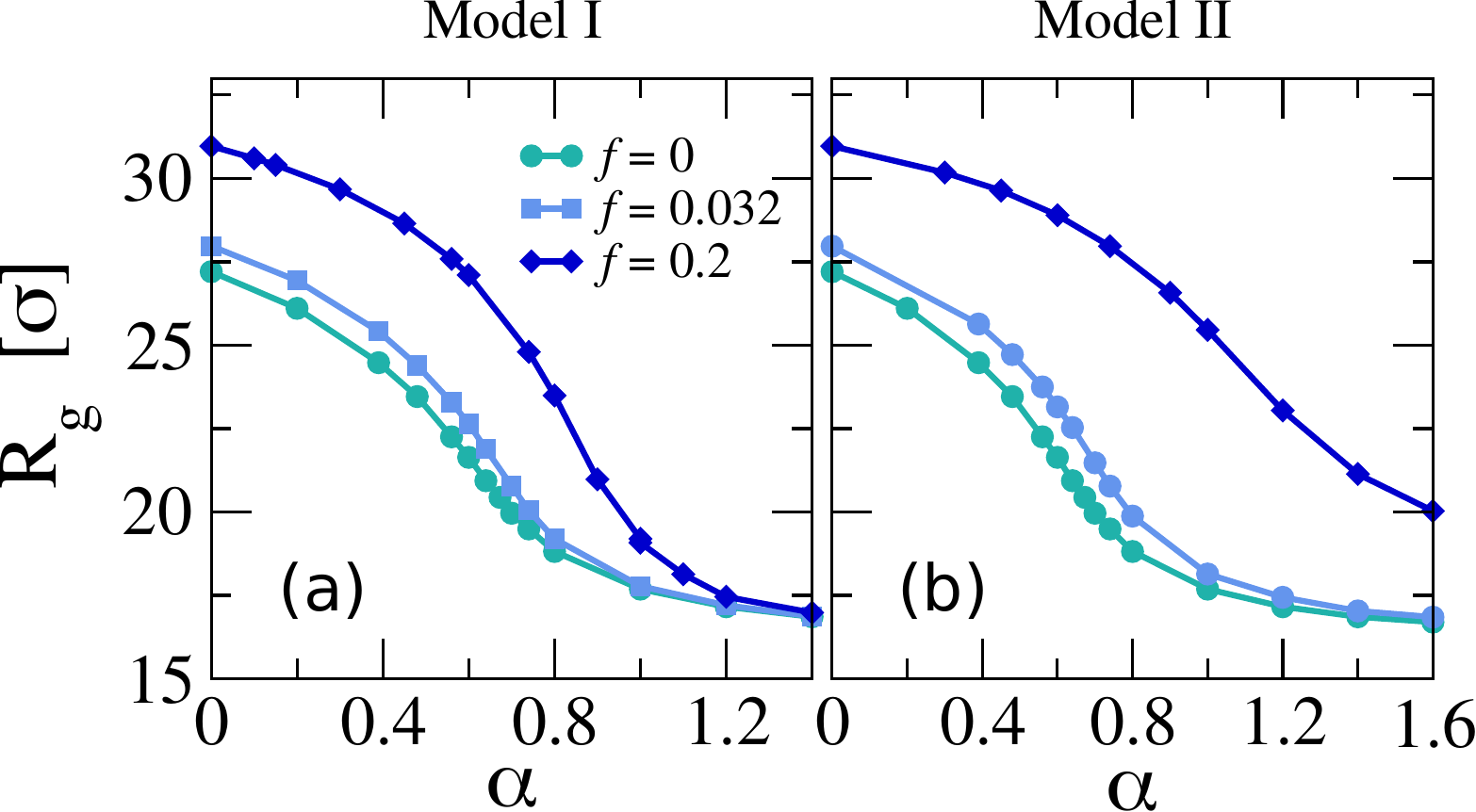}
\caption{\textbf{Swelling curves.} Radius of gyration $R_g$ as a function of the solvophobic parameter $\alpha$ and different values of $f$ for microgels with $N\approx 42000$ for the case where charged monomers (a) have a varying affinity for the solvent (\textit{Model I}) and (b) have always a good affinity for the solvent (\textit{Model II}). }
\label{fig:swelling42k}
\end{figure}
\begin{figure*}[h!]
\centering
\includegraphics[scale=0.59]{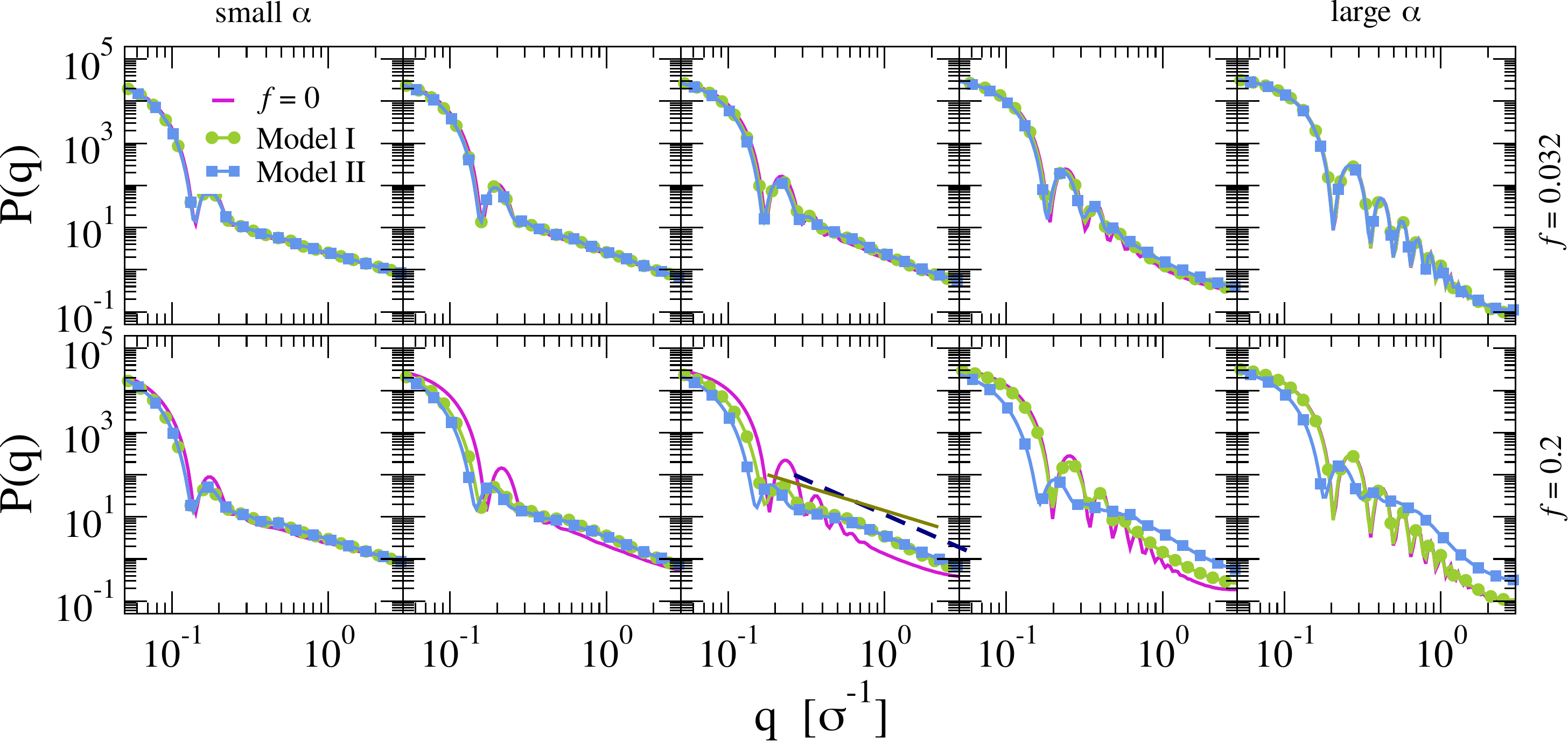}
\caption{\textbf{Form factors.} Form factors for charged microgels with $N\approx 42000$  and (top) $f=0.032$ and (bottom) $f=0.2$, simulated in implicit solvent for \textit{Models I} and \textit{II}. The models are compared at the same $\alpha$: for $f=0.032$, $\alpha = 0, 0.48, 0.64, 0.8, 1.4$; for $f=0.2$, $\alpha = 0, 0.6, 0.74, 1.0, 1.4$ (from left to right). The corresponding neutral case ($f=0$) is also displayed for comparison. Straight lines in the central panel of the bottom row  highlight the two power-law behaviours of the form factors at intermediate (full line) and high (dashed line) $q$ values, that are present for both models, extensively discussed in Ref.~\protect\cite{del2019numerical}.
}
\label{fig:formfactors42k}
\end{figure*}

We start by discussing the influence of charges on the VPT for microgels with $N\approx 42000$ in implicit solvent. As explained in the Methods section, we directly compare the case where the affinity of charged beads for the solvent varies (\textit{Model I}) to the case where it remains unchanged (\textit{Model II}).  Representative simulation snapshots  of the two models for the highest value of charge fraction investigated in this work ($f=0.2$) are reported in Fig.~\ref{fig:snaps42k}. Here we  focus on different swelling stages of the microgels upon increasing $\alpha$. We notice immediately that a large amount of inhomogeneities persists in \textit{Model II} at large $\alpha$, in contrast with the behavior of \textit{Model I} where a full collapse is achieved. This can be better quantified by the swelling curves, reporting the variation of the radius of gyration $R_g$ versus the effective temperature $\alpha$, that are shown in Fig.~\ref{fig:swelling42k} for different values of the charge fraction $f$. 
For both models we observe that the increase of  $f$ shifts the transition towards larger effective temperatures, but important differences arise at large $\alpha$, as displayed in the snapshots. In \textit{Model I}, where charged beads experience Coulomb as well as solvophobic interactions, the VPT occurs at all studied $f$, as shown in Fig.~\ref{fig:swelling42k}(a). Using the $\alpha$-temperature mapping established in Ref.~\cite{ninarello2019modeling} through a comparison to experiments, the VPT temperature observed for $f=0.2$ microgels would correspond to $T \approx 38^{\circ}$C.  However, experiments on ionic microgels, for which the amount of charges was systematically varied~\cite{holmqvist2012structure,capriles2008coupled}, have shown that even for values of $f$ smaller than $0.2$, the microgel does not collapse below $40^{\circ}$C. 

As hypothesized in our earlier work~\cite{del2019numerical}, \textit{Model I} neglects the interplay between the hydrophilic character of the co-polymer and its charge content. However, charged or polar groups, such as AAc groups, are known to remain hydrophilic even at high temperatures~\cite{wiehemeier2019synthesis},  which would increase the stability of the microgel in the swollen state with increasing $f$. We thus incorporate such a feature in \textit{Model II} by removing solvophobic interactions for charged microgel beads. The resulting swelling curves, shown in Fig.~\ref{fig:swelling42k}(b), clearly demonstrate that for $f=0.20$ the VPT is not encountered within the investigated solvophobicity range, up to values of $\alpha$ that would correspond to temperatures above $50^{\circ}$C, in qualitative agreement with experimental observations~\cite{holmqvist2012structure,capriles2008coupled,brandel2017microphase}.

\subsubsection{Structural properties}

It is now important to compare the two models from the structural point of view, to check whether major
differences arise.  We start the analysis by looking at the form factors which, in our previous work on \textit{Model I}~\cite{del2019numerical}, were shown to exhibit novel features with respect to neutral microgels. In particular, we found evidence  that for  $\alpha< \alpha_{\scriptscriptstyle \mathit{VPT}}$, the standard fuzzy-sphere-like model was not able to describe the numerical form factors, Instead, the emergence of two distinct power-law behaviours was found immediately after the first peak, at intermediate and high $q$ values, respectively~\cite{del2019numerical}. This was attributed to the presence of charges in the inhomogeneous structure of the microgel, which gives rise to different features for core and corona regions, each being characterized by a different domain size. It is now crucial to verify whether such distinctive behaviour also persists when the interactions among charged beads are modelled more realistically.

\begin{figure*}[h!]
\centering
\includegraphics[scale=0.59]{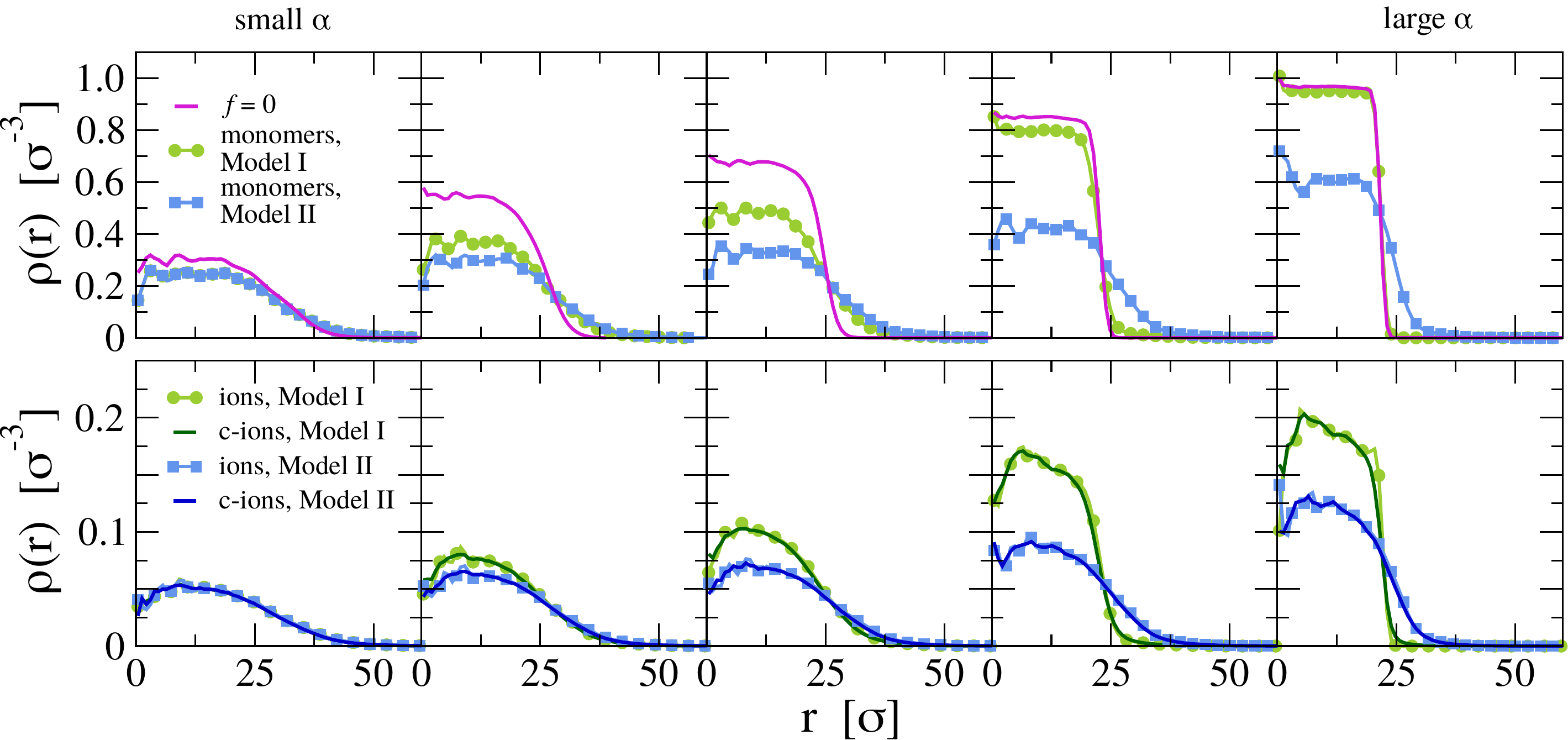}
\caption{\textbf{Density profiles.} Top panels show the monomers density profiles for an ionic microgel with $f=0.2$ and $N\approx 42000$ as a function of the distance from the microgel center of mass $r$ obtained in implicit solvent for \textit{Models I} and \textit{II}. Bottom panels report the ions and counterions (c-ions) density profiles for $f=0.2$ for both models. The models are compared at the same $\alpha$: $0, 0.6, 0.74, 1.0, 1.4$ (from left to right). The corresponding neutral case ($f=0$) is also displayed for comparison.
}
\label{fig:densprof42k}
\end{figure*}
\begin{figure}[t!]
\centering
\includegraphics[scale=0.38]{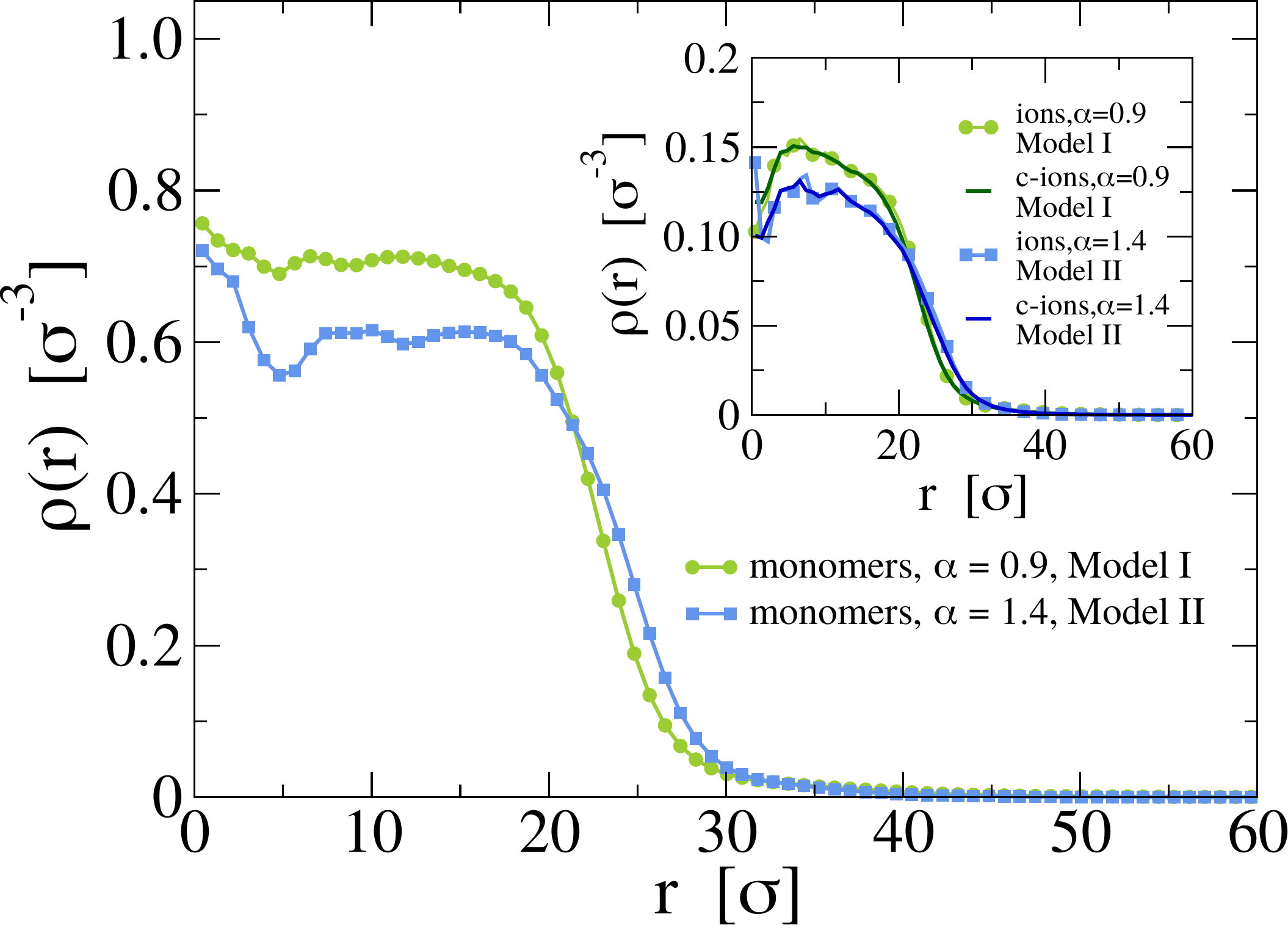}
\caption{\textbf{Comparison of \textit{Models I} and \textit{II} at the same $R_g$.} Radial density profiles for an ionic microgel with $f=0.2$ and $N\approx 42000$  at $R_g\approx 21$, where $\alpha=0.9$ and $\alpha=1.4$ for \textit{Models I} and \textit{II}, respectively. The inset shows the corresponding ions and counterions (c-ions) density profiles.
}
\label{fig:samerg42k}
\end{figure}
\begin{figure*}[h]
\centering
\includegraphics[scale=0.55]{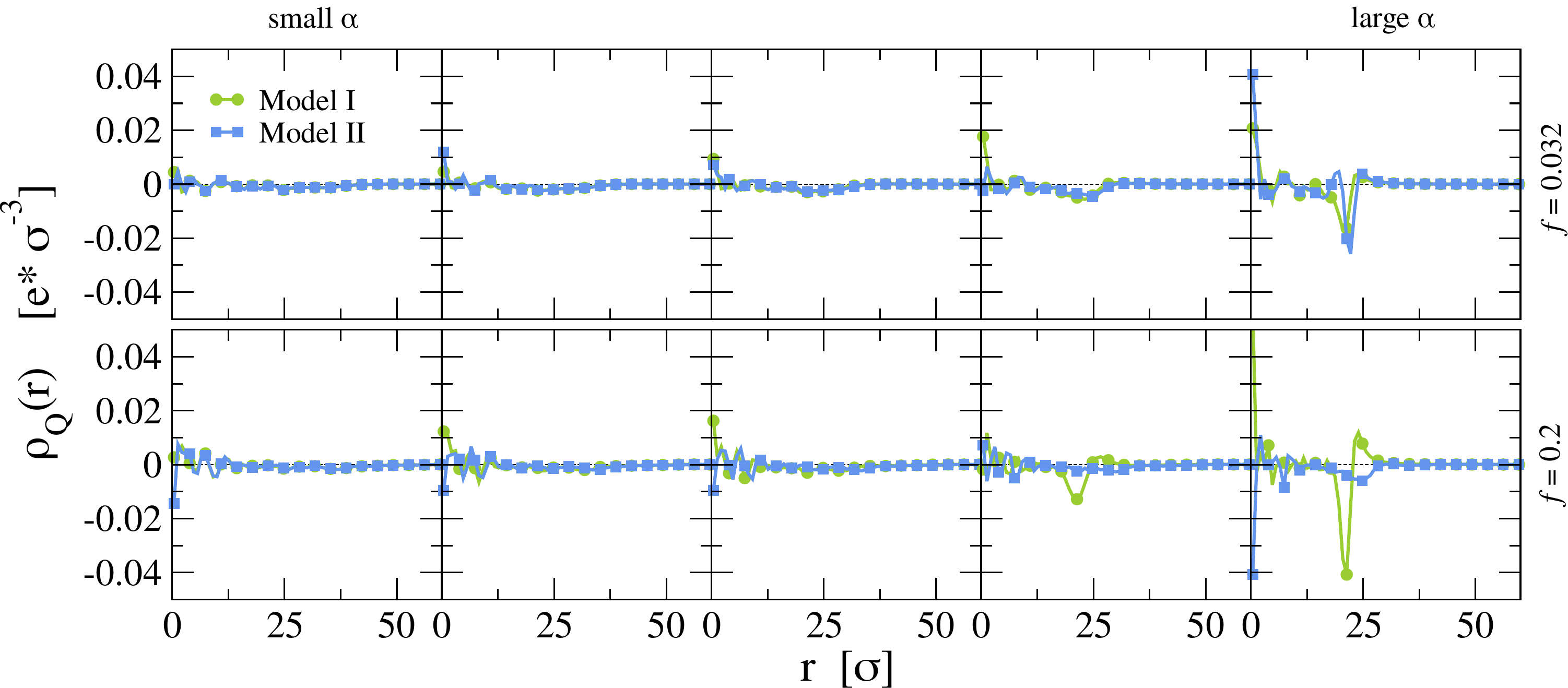}
\caption{\textbf{Charge density profile.} Net charge density profile $\rho_Q(r)$ as defined in Eq.~\protect\ref{eq:rho_charge} 
	for  ionic microgels with $N\approx 42000$  and (top) $f=0.032$ , (bottom) $f=0.2$,  as a function of the distance from the microgel center of mass $r$, simulated in implicit solvent for \textit{Models I} and \textit{II}. The models are compared at the same $\alpha$: for $f=0.032$, $\alpha = 0, 0.48, 0.64, 0.8, 1.4$; for $f=0.2$, $\alpha = 0, 0.6, 0.74, 1.0, 1.4$ (from left to right panels).
}
\label{fig:chargeprof42k}
\end{figure*}

Fig.~\ref{fig:formfactors42k} reports the form factors for \textit{Models I} and \textit{II} with $f=0.032$ and $f=0.2$, in comparison to the neutral case ($f=0$), at different values of $\alpha$.
For $f=0.032$, the amount of charges in the microgel is still too low to observe differences between the two models and the neutral microgel. Also, at large $\alpha$, the form factor is that of a collapsed microgel in all cases, as expected from the swelling curves in Fig.~\ref{fig:swelling42k}.
For $f=0.2$ and low enough $\alpha$, the behaviour of the two models is again very similar, with the form factors of ionic microgels showing a first peak that is systematically smaller with respect to that of the neutral case. At intermediate $\alpha$, we find that two power-law-like behaviours are compatible with both sets of data  for charged microgels, while the neutral case does not show such a feature. This finding, already elaborated in Ref.~\cite{del2019numerical}, appears to be a distinctive feature of our numerical model of ionic microgels and is the result of the combination of a random charge distribution within a disordered, heterogeneous network topology with the explicit treatment of ions and counterions.  Such a distinctive feature was tentatively attributed to the different degree of swelling of the corona and of the core, but still awaits a direct experimental confirmation. However, hints of a similar two-step decay for $P(q)$ were reported in Ref.~\cite{fernandez2002structural} and would certainly deserve further investigation in future experiments.
\\
On the other hand, major differences between the two charged models arise for large values of $\alpha$. Indeed, in \textit{Model I} the microgel approaches
and crosses the VPT leading to a fully collapsed state, while in \textit{Model II} it remains in a quasi-swollen configuration for all studied $\alpha$. Consequently, for high $\alpha$ values, the form factor does not resemble that of a homogeneous sphere, with only a second peak becoming evident, as opposed to the neutral case where many sharp peaks emerge. We notice that \textit{Model I} fully coincides with the neutral case for very large $\alpha$, even for $f=0.2$.  

In Fig.~\ref{fig:densprof42k}, we compare the monomers density profiles for the two models as a function of $\alpha$. These data further indicate that, for \textit{Model II}, the microgel does not achieve a collapsed state, as also visible from the behaviour of the profiles of charged monomers and of counterions, respectively. These are reported in the bottom panels of Fig.~\ref{fig:densprof42k}, showing that, for both models, the counterions are always found to be very close to the charged monomers, in order to neutralize the overall charge of the microgel. However, all profiles remain much more extended for \textit{Model II} as compared to \textit{Model I}, for all $\alpha$. We stress that the comparison is performed for microgels with different affinity of the charged monomers for the solvent at the same $\alpha$, which corresponds to very different swelling conditions, as evident from Fig.~\ref{fig:swelling42k}.  Additional information can be extracted by comparing the two cases for a similar value of $R_g$, as reported in Fig.~\ref{fig:samerg42k}. Also in this case, we  find that \textit{Model II} displays a more slowly decaying radial profile, albeit having a very similar mass distribution with respect to \textit{Model I}, which is  due to the presence of more stretched external dangling chains.
Similar results also apply to ions and counterions profiles, that are shown in the inset of Fig.~\ref{fig:samerg42k}: even at the same $R_g$, there is a surplus of charges at the surface in the case where the affinity of charged monomers for the solvent does not change with the effective temperature (\textit{Model II}).
Overall, these findings confirm an enhanced stabilization of the swollen configuration operated by the charged groups of the microgel, hindering the tendency of the remaining (neutral) monomers to collapse.

To complete the structural analysis of the two models, it is instructive to consider the net charge density profile inside the microgels, that is reported in Fig.~\ref{fig:chargeprof42k} for both $f=0.032$ and $f=0.2$. We confirm that, for both models, the net charge of the core region is roughly zero. However, it was shown in Ref.~\cite{del2019numerical} that in the collapsed configuration a charged double layer  arises at the surface of microgels, signalling the onset of a charge imbalance that grows with $\alpha$. This feature, that is clearly visible in the behaviour of \textit{Model I} at high $\alpha$ for all values of $f$,  is also present for \textit{Model II} for the low charge case ($f=0.032$). However, the double peak in the net charge distribution is smeared out for $f=0.2$, due to the fact that, up to the largest explored values of $\alpha$, the microgel does not fully collapse. In this way,  it maintains a low concentration of charged beads, that is always roughly balanced by counterions, resulting in a rather uniform charge profile. Instead, the peaks at the surface appear when the microgel collapses: this is indeed the case for both models at low charge fraction and even for large $f$ when charged monomers are assigned a solvophobic behaviour (\textit{Model I}).

We conclude from this analysis that the hydrophilicity of the charged monomers at all effective temperatures enhances the tendency of the microgel to remain swollen, even when most of the monomers experience a very large solvophobic attraction. Thanks to the charge neutralization operated by counterions, the microgel remains very stable in a rather swollen configuration up to very large $\alpha$, avoiding collapse for large enough values of $f$. This scenario agrees well with experimental observations, where the suppression of the VPT~\cite{capriles2008coupled,brandel2017microphase,wiehemeier2019synthesis} is found when the concentration of charged hydrophilic groups in the polymer network is large enough. These considerations imply that \textit{Model I} should not be used to describe microgels with high charge content. Indeed, its identical treatment of the solvophilic character of both neutral and charged monomers leads the particle to collapse at extremely high $\alpha$. Incidentally, we report that this was observed also for unrealistic values of $f$ up to $0.4$ (not shown), in evident contrast with experiments. We will thus rely on  \textit{Model II} in the future to correctly incorporate charge effects in modelling microgels in a realistic fashion.

%=============================================================
%=============================================================
%=============================================================

\subsection{\textbf{Solvent effects}}

We now go one step further in modelling ionic microgels, by explicitly adding the solvent to the simulations.
This is a necessary prerequisite to tackle phenomena that cannot be described with an implicit solvent, e.g. situations in which hydrodynamics or surface tension effects at a liquid-liquid interface~\cite{camerin2019microgels} play a fundamental role.
In this subsection, we compare results for swelling behaviour and structural properties of the microgels for implicit and explicit solvent simulations. In particular, we restrict our discussion to \textit{Model II}, having established this to be more in line with experimental observations. Since simulations with an explicit solvent require a much higher computational effort, we limit the following discussion to microgels with $N\approx 5000$.

\subsubsection{Swelling curves and explicit-implicit ($a$-$\alpha$) mapping}
We start by reporting the swelling curves of charged microgels, stressing the point that they have been obtained by fixing the value of $a_{ch,s}$, which tunes the solvophilic properties of charged beads and counterions. We find that setting $a_{ch,s} = 0$, while $a_{m,s}\equiv a$ varies, the explicit model is essentially equivalent to the implicit one. This means that it is possible to find a relation that links every implicit system with a certain value of the solvophobic attraction $\alpha$ to an explicit one with solvophobic parameter $a$ that shows the same structure and swelling properties.

\begin{figure*}[t!]
	\centering
	\includegraphics[scale=0.54]{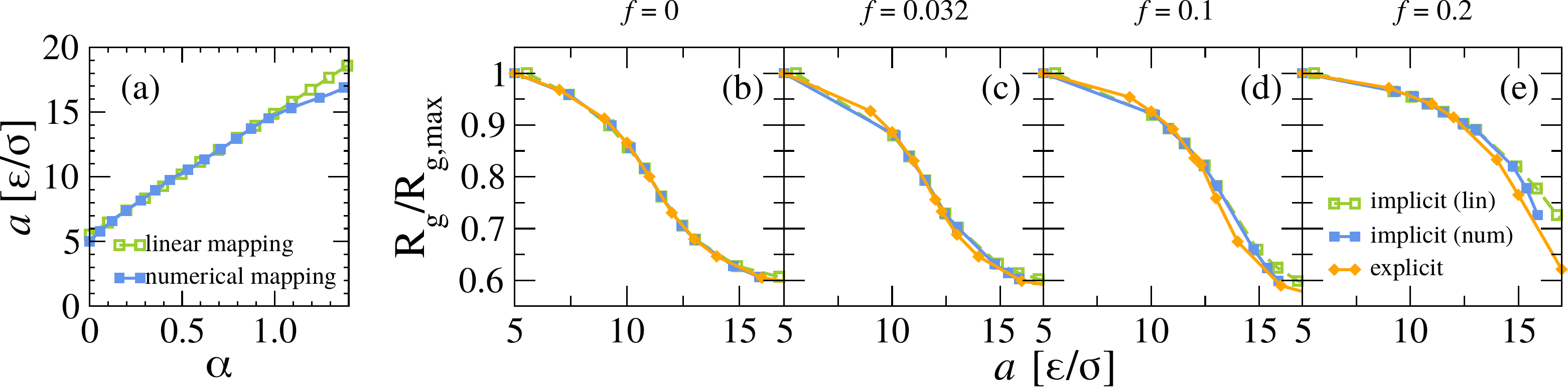}
	\caption{\textbf{Implicit-explicit solvent mapping and swelling curves.} 
		(a) Mapping between $\alpha$ and $a$ obtained by comparing neutral microgels with implicit (\textit{Model II}) and explicit solvents: the linear mapping is expressed by Eq.~\protect\ref{eq:lin_mapping} and the numerical mapping via Eq.~\protect\ref{eq:num_mapping}; 
		(b-e) Normalized radius of gyration $R_g/R_{g,max}$ as a function of the swelling parameter $a$ for microgels with different charge content: (b) neutral, (c) $f=0.032$, (d) $f=0.1$ and (e) $f=0.2$, for explicit (full lines and filled diamonds) and implicit  solvent conditions (rescaled along the horizontal axis using the linear mapping $a_{\text{lin}}(\alpha)$, dashed lines and empty squares, and using the numerical mapping  $a_{\text{num}}(\alpha)$, full lines and filled squares). The present figure and the following ones refer to the same microgel topology with $N \approx 5000$.		}
	\label{fig:swellingcurves_mapping}
\end{figure*}

In order to establish such a $a$-$\alpha$ mapping, we explored two different routes. The first one, referred to as linear mapping in the following, is based on the assumption that the dependence of $a$ on $\alpha$ is linear, as previously adopted for neutral microgels \cite{camerin2018modelling}. In this way, the mapping relation is  obtained through a horizontal rescaling of the relative swelling curves $R_g^{\text{imp}}(\alpha)/R_g^{\text{imp}}(\alpha = 0)$ and $R_g^{\text{exp}}(a)/R_g^{\text{exp}}(a = 0)$ for the neutral implicit and explicit microgels onto each other. Specifically, given two points for each curve, $(a_1 , a_2)$ and ($\alpha_1 , \alpha_2$), the rescaled $x$-coordinate is calculated using the following relationship:
\begin{equation}\label{eq:lin_mapping}
	a_{lin}(\alpha)=\left(\alpha-\langle \alpha \rangle\right)\Delta a/\Delta \alpha  + \langle a \rangle
\end{equation}
where $\langle x \rangle=0.5(x_1+x_2)$ and $\Delta x= x_1-x_2$ with $x=a,\alpha$.
The second mapping $a_{\text{num}}(\alpha)$, referred to as numerical mapping, has been obtained by numerically inverting the equation
\begin{equation}\label{eq:num_mapping}
	R_g^{\text{imp}}(\alpha)/R_g^{\text{imp}}(\alpha = 0) = R_g^{\text{exp}}(a)/R_g^{\text{exp}}(a = 0),
\end{equation}
after spline fitting the two swelling curves.
We report both mapping relations in Fig. \ref{fig:swellingcurves_mapping}(a), finding that they fall onto each other for almost the entire range of investigated solvophobic parameters in the two models, confirming the overall correctness of the assumption of linearity. However, we find some differences in the region $\alpha > 1.0$ ($a > 15$).  Having established the mapping for neutral microgels, we now use it to directly remap also the results for ionic microgels for all studied $f$ without any further adjustments.

\subsubsection{Swelling behaviour}
The normalized swelling curves with varying charge fraction $f$, comparing implicit and explicit solvent, are reported in Fig.~\ref{fig:swellingcurves_mapping}(b-e). Data from implicit simulations are mapped via both methods described above. For the neutral case, the presence of the solvent does not affect the swelling behaviour, as shown in Fig.~\ref{fig:swellingcurves_mapping}(b), where no appreciable differences are found between linear and numerical mapping even at high $\alpha$. Using the same relations for comparing charged microgels in explicit and implicit solvent, we find that, remarkably, the same swelling behaviour works for all charge contents. The swelling curves are virtually identical, which ensures that the inclusion of the solvent does not alter the microgel behavior in temperature even in the presence of charges. Small deviations, as expected from Fig.~\ref{fig:swellingcurves_mapping}(a),  appear only at large $\alpha$ values, being more pronounced for high charge content. This confirms the robustness of the DPD model which, as already discussed in Ref.~\cite{camerin2018modelling}, does not induce spurious effects, e.g. due to excluded volume, even in the collapsed state. An important result of this work is that, even in the presence of an explicit solvent, the microgel at high $f$ does not fully collapse at large $\alpha$, being entirely equivalent to implicit \textit{Model II} and compatible with experimental findings.

\begin{figure*}[t!]
	\centering
	\includegraphics[scale=0.55]{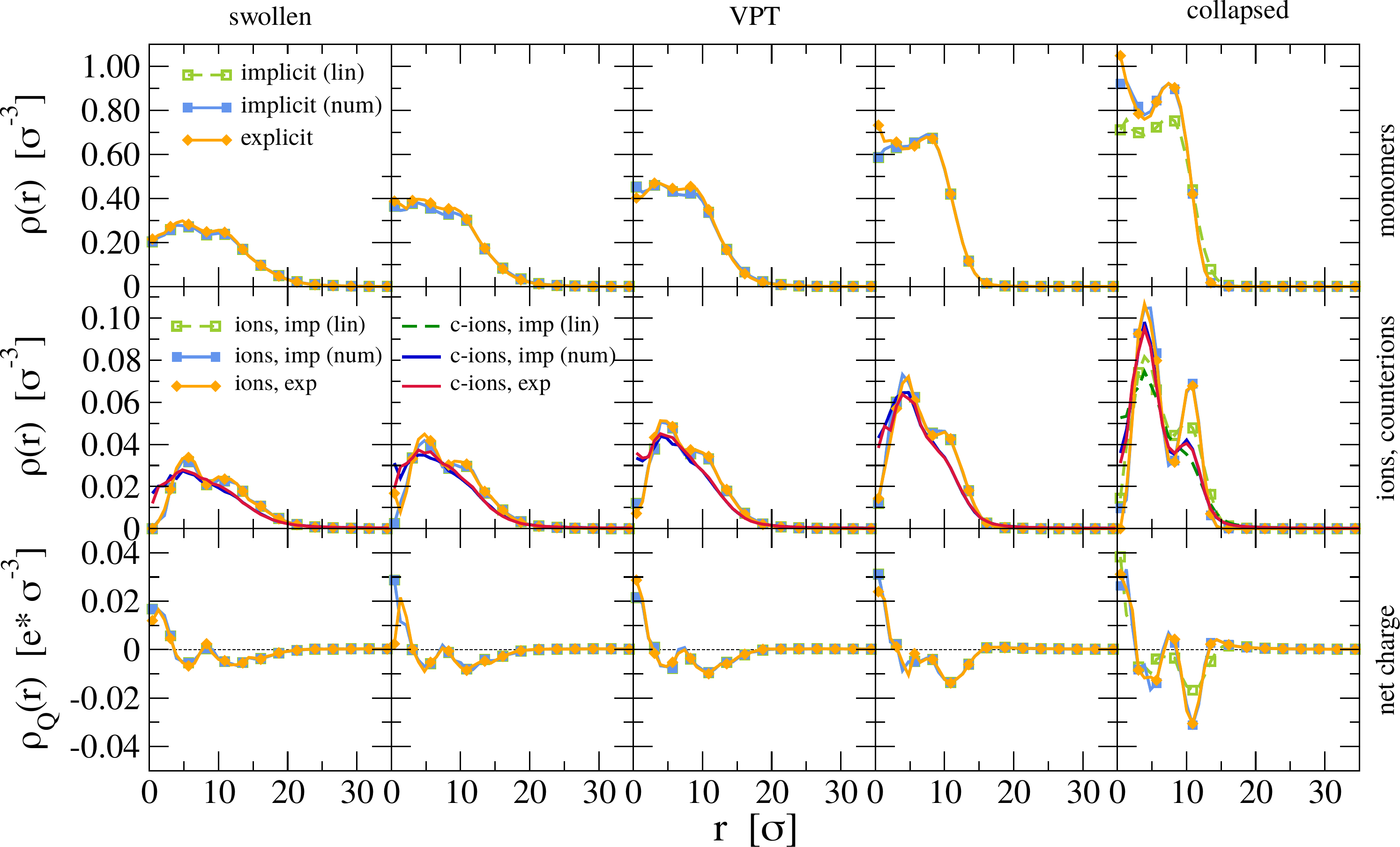}
	\caption{\textbf{Density profiles.} Density profiles of monomers (top row), charged beads and counterions (middle row) and net charge (bottom row) for ionic microgels with $f=0.1$ as a function of the distance from the microgel center of mass $r$ obtained in explicit and implicit solvent conditions. Curves from the explicit case refer to values of $a = 5, 11, 12.3, 14, 16$,	from the (left) swollen to the (right) collapsed state.
	Implicit and explicit solvent cases are compared at values of $\alpha$ approximately corresponding to each $a$ value according to both the linear ($\alpha = 0, 0.56, 0.74, 1.0, 1.1$) and the numerical ($\alpha = 0, 0.56, 0.74, 1.0, 1.2$) mapping. 	\label{fig:densprof_f0.1}
}
\end{figure*}

\subsubsection{Structural properties}

In this subsection, we will show that  the implicit and explicit solvent treatments with the newly established numerical mapping (Eq.~\ref{eq:num_mapping}) lead to identical structural features of the microgels. Small differences arise when using the linear mapping (Eq.~\ref{eq:lin_mapping}) at high $f$ and large values of $\alpha$. 

We show in Fig.~\ref{fig:densprof_f0.1} the monomer (top panels), ion and counterion (middle panels) and charge (bottom panels) density profiles only for the $f=0.1$ case, since similar results are also found for the other studied charge fractions. Reported data for different values of monomer-solvent interactions show an overall similarity between  implicit and explicit solvent descriptions
at all swelling conditions. Small deviations arise only for $f=0.2$ for states with the highest values of $a$ or $\alpha$, when using the linear mapping: as we can observe from the rightmost panels of Fig.~\ref{fig:densprof_f0.1}, the linear mapping fails to associate implicit and explicit states in the most collapsed state, where a visible difference arises between the profiles.

The distribution of ions and counterions within the microgel is an  observable that should be more sensitive to the presence of the solvent. However, quite remarkably, also in this case, we find excellent agreement between the two models, as shown in the middle panels of Fig.~\ref{fig:densprof_f0.1}. In particular, the emergence of a clear double-peak structure in the ion distribution is found in both models for large $\alpha$ (implicit) and $a$ (explicit), signalling an accumulation of ions at the exterior surface of the microgels. This can be understood from the fact that ions, remaining always hydrophilic, never completely collapse onto the core of the particle. Thus, the appearance of a peak at distances corresponding to the outer region of the microgel is the result of an attempt of ions to maximize their contact with solvent. This is preceded by a minimum, which indicates a region where ions are depleted within the particle.

This feature is the echo of the minimum that arises in the net charge density distributions,  already anticipated for the large microgel treated with the implicit model in Fig.~\ref{fig:chargeprof42k}. Importantly, a minimum also occurs in $\rho_Q(r)$ for smaller microgels, as shown in 
 the bottom panels of Fig.~\ref{fig:densprof_f0.1}, for the most collapsed conditions. Here a charged double layer is clearly present, with an excess of positive charges inside the microgel corona due to the increased amount of counterions in this region. At the same time, a negative charge surplus is found at the surface of the microgel, since charged ions preferably remain in contact with solvent particles. The net charge distribution is also identical for explicit and implicit solvent when using the numerical mapping, with again very small differences arising for the linear mapping at large $\alpha$.

\begin{figure*}[t!]
	\centering
	\includegraphics[scale=0.55]{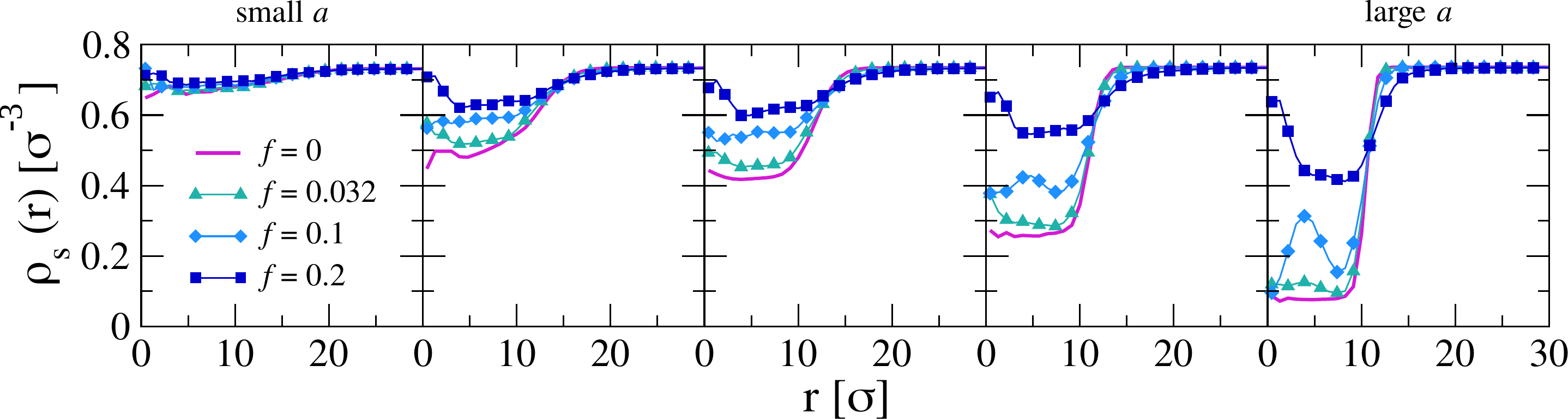}
	\caption{\textbf{Solvent density profiles} for charged microgels of different $f$ values, as a function of the distance from the microgel center of mass $r$. The different panels refer to $a = 5, 11, 12, 14, 16$ from (left) good to (right) bad solvent conditions. 	\label{fig:dens_solvent}
}
\end{figure*}

It is important to notice that, although a double layer was also observed with the implicit solvent in Ref.~\cite{del2019numerical} (equivalent to \textit{Model I}), the two distributions (the one in Fig.~\ref{fig:densprof_f0.1} of the current manuscript and that reported in Fig.~6 of Ref.~\cite{del2019numerical}) have opposite signs. Indeed in Ref.~\cite{del2019numerical} the superposition of electrostatic and solvophobic effects led to an accumulation of counterions at the microgel surface, with the onset of a seemingly Donnan equilibrium~\cite{hunterfoundations}. Notwithstanding the different origin of the double layer, both models demonstrate that an almost perfect neutrality is achieved within the core of the microgel, and it is only at the surface that inhomogeneous distributions appear. Besides, the reduced size of the microgels studied with the explicit solvent facilitates the onset of peaks due to the increased surface-to-volume ratio of the microgels.
A more precise assessment of size effects and a careful comparison to experiments will be the subject of future works.

Finally, the explicit solvent model allows us to quantify the amount of solvent that is located inside the microgel as temperature increases. This is 
illustrated in Fig.~\ref{fig:dens_solvent}, where the solvent density profile $\rho_s(r)$ is reported for different values of $a$ and all investigated charge fractions. These plots confirm the reduced tendency to collapse of charged microgels which retain a large amount of solvent within the network structure. No inhomogeneities within the microgel are in general observed. At large $f$ and $\alpha$ some oscillations arise which may be due to reduced statistics. Finally, this study confirms that even at temperatures above the VPT there is quite a residual amount of solvent within the microgel, that is significantly enhanced by increasing the charge. These findings are in line with expectations~\cite{brandel2017microphase,bischofberger2015new}, that are thus confirmed by our simulations.

\section{Conclusions}

In this work we report an extensive numerical study of single microgel particles, a prototype of soft colloids that is of great interest for the colloidal community, particularly for the formation of arrested states with tunable rheological properties~\cite{vlassopoulos2014tunable}, including glasses~\cite{mattsson2009soft,philippe2018glass} and gels~\cite{wu2003interparticle}. The use of different polymers within the microgel network allows to exploit responsiveness to different control parameters, such as temperature and pH, giving rise also to unusual responses in the fragility of the system~\cite{mattsson2009soft,nigro2017dynamical,gnan2019microscopic}. 

In order to be able to model dense suspensions of these soft particles, we can rely on two possible strategies. On one hand, we can exploit highly coarse-grained models, such as the Hertzian one, which completely neglect the polymeric degrees of freedom of the particles and thus cannot reproduce the complex phenomenology observed in experiments in the gel or glassy regimes, such as shrinking, faceting and interpenetration~\cite{conley2017jamming,conley2019relationship}. On the other hand, we can try to model  a single microgel in a realistic way, aiming to reproduce its structural properties and, from this, to build effective interactions which retain the polymeric features of the single particle. 

Adopting the second strategy, the aim of this work is to improve the current numerical modelling of single ionic microgels with randomly distributed charged groups, aiming to describe PNIPAM-co-PAAc microgels across the Volume Phase Transition. In particular, we assess two different ways to model the interactions of the charged monomers belonging to the polymer network, either considering or not a solvophobic attraction that mimics their hydrophilic/hydrophobic interactions. We find that, as long as the charged groups maintain the same affinity for the solvent, the tendency of the microgel to remain in swollen conditions is enhanced even at high effective temperatures. Thus, for a charge fraction of $f=0.2$ we find no evidence of the collapse of the microgel within the investigated range of our simulations, in agreement with experimental observations that are currently available~\cite{holmqvist2012structure,capriles2008coupled,brandel2017microphase,wiehemeier2019synthesis}. 
This result is different from the case where charged beads also attract each other like neutral monomers upon increasing temperature, which undergoes a Volume Phase Transition to a fully collapsed state~\cite{del2019numerical}. Despite this fundamental difference, the structural properties of the microgels treated with both models are rather similar, especially at low and intermediate temperatures. For instance, we confirm that the peculiar power-law regimes observed in the form factors are independent of the chosen model.

Having established the most appropriate modelling for charged monomers, we then performed another necessary step in the modelling of ionic microgels, namely to explicitly consider the presence of the solvent, which may affect the rearrangement of the charges during the swelling-deswelling transition. To this aim, we build on previous results showing that for neutral microgels a description with an explicit solvent can be directly and quantitatively superimposed to the implicit modelling by using a DPD representation of the solvent, leaving unchanged the treatment of the polymer network with a bead-spring model. In this way, the solvophobic potential in Eq.~\ref{eq:valpha}, modulated by the parameter $\alpha$, is replaced by the DPD repulsive interactions between monomers and solvent. The latter is varied through a change of the parameter $a$ controlling the repulsion between non-charged beads, while the interaction between charged monomers always retains a solvophilic nature.

We have thus carried out a careful comparison between explicit and implicit solvent treatments, finding quantitative agreement between the two.
Interestingly, the relation among $a$ and $\alpha$ established by the comparison of neutral microgels can be used also to compare charged microgels, even with large values of $f$ (some deviations occur only at $f=0.2$ and large $\alpha , a$ values), where the same correspondence between implicit and explicit solvent states is retrieved. We showed that a linear mapping between the two control parameters of the interactions in the implicit and explicit case is sufficient to obtain a very good agreement between the two descriptions.

From our analysis of the internal structure of the microgels across the VPT, we found that counterions have a rather similar distribution within the microgel core, effectively neutralizing the internal charge at small distances, but being in excess close to the surface. This gives rise to a charged double layer for large values of $a$ and $\alpha$. Interestingly, such peaks in the charge density distributions are swapped with respect to the case of \textit{Model I}, where ions do not experience a tendency to remain at the surface, since they are also treated as solvophobic. These detailed predictions will have to be compared to experiments on ionic microgels as a function of charge fraction, pH and $T$, in order to establish the limit of validity of our model and to further improve it, towards a more realistic description of experimental microgels.

In perspective, this work paves the way to study realistic charged microgels in diffusing conditions, such as in electrophoresis and thermophoresis experiments~\cite{wongsuwarn2012giant}, or at liquid-liquid interfaces and to calculate their effective interactions, similarly to what has been done for neutral microgels~\cite{camerin2019microgels,camerin2020microgels}. In this way, we will be able to determine the conditions under which electrostatic effects play a dominant role over elastic ones. Another important line of research will be the assessment of the role of the network topology: examples of interesting cases whose properties could be investigated are microgels consisting of two interpenetrated networks~\cite{nigro2017dynamical,nigro2019study} or ultra-low crosslinked~\cite{bachman2015ultrasoft,scotti2020flow} and  hollow~\cite{nayak2005hollow,nickel2019anisotropic} microgels. Finally, we hope that our theoretical efforts will stimulate further experimental activity on charged microgels to verify the predicted behaviour so that it will be possible to tackle the investigation of dense suspensions in the near future.
 
\section*{Acknowledgments}
This research has been performed within the PhD program in "Mathematical Models for Engineering, Electromagnetics and Nanosciences". We acknowledge financial support from the European Research Council (ERC Consolidator Grant 681597, MIMIC). FC and EZ also acknowledge funding from Regione Lazio, through L.R. 13/08 (Progetto Gruppo di Ricerca GELARTE, n.prot.85-2017-15290).

\section*{References}
\bibliography{jpcmbib}

\end{document}